\documentclass[a4paper,11pt]{article}
\pdfoutput=1 

\usepackage{jcappub} 

\usepackage[T1]{fontenc} 
\usepackage{color,ulem}

\newcommand{\lta}{\lower 2pt \hbox{$\, \buildrel {\scriptstyle <}\over {\scriptstyle \sim}\,$}}
\newcommand{\gta}{\lower 2pt \hbox{$\, \buildrel {\scriptstyle >}\over {\scriptstyle \sim}\,$}}
\definecolor{blazeorange}{rgb}{1.0, 0.4, 0.0}
\definecolor{seagreen}{rgb}{0.18, 0.55, 0.34}
\definecolor{rufous}{rgb}{0.66, 0.11, 0.03}
\definecolor{royalfuchsia}{rgb}{0.79, 0.17, 0.57}
\definecolor{scarlet}{rgb}{1.0, 0.13, 0.0}
\definecolor{royalpurple}{rgb}{0.47, 0.32, 0.66}

\def\apjl{ApJL}
\def\apj{ApJ}
\def\mnras{M.N.R.A.S.}
\def\aap{A\&A}

\def\araa{Ann. Rev. A\&A}

\def\prd{Phys. Rev. D.}

\def\ssr{Space Sci. Review}
\def\prl{Phys. Rev. Lett. }
\def\nat{Nature}
\def\nar{Nature Astronomy}

\title{\boldmath Ultra High Energy Cosmic Rays from Tidal Disruption Events}

\author[a,1]{T. Piran,\note{Corresponding author.}}
\author[b,c,d]{and P. Beniamini}

\affiliation[a]{Racah Institute for Physics, The Hebrew University, Jerusalem, 91904, Israel}
\affiliation[b]{Department of Natural Sciences, The Open University of Israel, P.O Box 808, Ra'anana 4353701, Israel }
\affiliation[c]{Astrophysics Research Center of the Open University (ARCO), The Open University of Israel, P.O Box 808, Ra'anana 4353701, Israel }
\affiliation[d]{Department of Physics, The George Washington University, 725 21st Street NW, Washington, DC 20052, USA}
\emailAdd{tsvi.piran@mail.huji.ac.il}
\emailAdd{pazb@openu.ac.il}

\abstract{The tidal disruption event AT2018hyz, was a regular optically detected one with no special prompt features. However, almost three years after the disruption it suddenly displayed a fast-rising radio flare. The flare is most naturally interpreted as arising from an off-axis relativistic jet. We didn't see the jet at early times as its emission was relativistically beamed away from us. However, we could see the radiation once the jet has slowed down due to interaction with the surrounding matter. Analysis of the radio data enabled estimates of the jet's kinetic energy and its opening angle as well as the conditions (size and magnetic field) within the radio-emitting region. We show here that such a jet satisfies the Hillas condition for the acceleration of UHECRs to the highest energies. We also show that the rate and total power of this event are consistent with the observed luminosity density of UHECRs. These results strongly support earlier suggestions that TDEs are the sources of UHECRs.   }

\begin{document}
\maketitle
\flushbottom

\section{Introduction}
\label{sec:intro}
Ultra-high energy cosmic rays (UHECRs) with energies up to $10^{20}$eV were first detected more than sixty years ago. Still, their sources of are unknown, making them one of the most interesting astronomical puzzles. { To reach such a high energy, the Hillas criteria \cite{Hillas1984} states that the particles' acceleration time should be shorter than their escape time. For that to hold  the size system and its magnetic field should be large (see Eq \ref{eq:Hillas} below). This implies, in turn, } that the sources must be very luminous. At the same time photopion production of UHECRs protons on the CMB or photodissociation of nuclei limit their propagation distances to about 100 Mpc (for the highest energy UHECRs). Thus the luminous sources must be local and easily identified. However,  deflection in the intergalactic and our own Galactic magnetic fields divert UHECRs from their original directions. This of course  complicates the identification of the sources. 
As there are no specific directions from which UHECRs accumulate  there is no single dominant source. A lack of correlation with any known population of luminous  sources suggests that the sources might be transients. Indeed two transient sources:  flaring Active Galactic Nuclei (AGNs) \cite{Farrar2009} and Gamma-ray Bursts (GRBs) \cite{Waxman1995,Vietri1995} have been proposed. 

GRBs  satisfy the Hillas criteria. However, their total power is small by at least one order of magnitude compared to the rate of production of UHECRs (e.g. \cite{Eichler2010}). 
Recent upper limit \cite{Abbasi2023} on the neutrino flux from the nearby and powerful GRB 221009A, put an independent strong limit on UHECR production by this source and GRBs in general \cite{Kruiswijk2023}.  Tidal disruption events (TDEs) are natural transients that take place typically around dormant super massive black holes in galactic centers.  A fraction of TDE eject relativistic jets. Already the first detected jetted TDE (Swift J1644) revealed that the shocks arising from the  interaction of the jet with the surrounding matter, a process resembling a GRB afterglow, satisfy the Hillas condition for UHECR acceleration \cite{Farrar2014}. { Subsequently, several authors \cite{Pfeffer2017,Zhang2017,Biehl2018,Guepin2018,Winter2023}  considered further the possibility that  TDEs are UHECR sources, discussing also the prospect that TDEs produce both UHECRs and high energy neutrinos.  }

We explore here recent developments in observations and theory of TDEs in this context. In particular, we consider the implication of  observations of some jetted TDEs and we explore again the question whether those relativistic jets  satisfy  
the Hillas criteria and whether the total power of TDEs is sufficient to supply the observed flux of UHECRs. We begin (in \ref{sec:obs}) with a very brief summary of observational features of UHECRs (\ref{sec:UHECRobs}) and of jetted TDEs (\ref{sec:jetsobs}). We continue (in \ref{sec:Hillas}) with a discussion of the Hillas criteria and  the conditions within shocks produced due to the interaction of TDE jets with the surrounding matter. We discuss (in \ref{sec:rates}) the rates of jetted TDEs and the implied luminosity density of these sources. We conclude in \ref{sec:conc}, summarizing our findings and possible future confirmation of these ideas.

\section{Observations}
\label{sec:obs}
\subsection{UHECRs }
\label{sec:UHECRobs}
We begin with a brief description of basic properties of UHECRs (see \cite{Globus2023} and references therein for further details). \hfill\break
$\bullet$ The UHECR energy  extends up to $\sim 10^{20}$ eV. The spectrum shows an ``ankle" at 5 EeV  and strong suppression is observed above 50 EeV. \hfill\break
$\bullet$ The UHECRs luminosity density above 5 EeV is $ {\cal L}_{\rm UHECR} = 6 \times 10^{44}$ erg Mpc$^{-3}$ yr$^{-1}$ \cite{PAOL2020}.\hfill\break 
$\bullet$ The composition above the ankle is dominated by nuclei. The nuclei become heavier as the energy increases and they are mostly CNO or heavier above $40$ EeV  \cite{Coleman2023}. \hfill\break
$\bullet$ The only significant departure from isotropy is the $6.9 \sigma$ dipole at energies above 8 EeV \cite{PAO_dipole2017}. At lower energy the observations are compatible with isotropy. No clear association with any population of sources have been identified \footnote{Note, however, that diffusion in the intergalactic and Galactic magnetic fields may divert significantly UHECRs trajectories away from the direction of their original source (see e.g. \cite{Farrar2000,Allard2012,Harari2014,Globus2017a,Globus2019,Ding2021})}.

\subsection{Jetted TDEs}
\label{sec:jetsobs}
Out of about a hundred TDEs, several  have been identified as harboring relativistic jets \cite{DeColle2020}.  
The identification was generally made, following the first TDE with an identified jet, Swift J1644, by a prompt non-thermal
prompt non-thermal X-rays \cite{Burrows2011,Bloom2011}. Likely due to obscuration, optical emission was not observed in Sw J1644.  Radio emission, detected within a few days \cite{Zauderer2011}  provided further support for a relativistic jet. Sw J2058 \cite{Cenko2012} and Sw J1112 \cite{Brown2015} are two other jetted TDEs detected by Swift. Multiwavelength emission, including non-thermal X-rays was observed from AT2022cmc \cite{Andreoni2022}, suggesting that it also included a jet. Recently, Matsumoto and Piran \cite{Matsumoto2023} (see also \cite{Sfaradi2023} proposed a new way to identify off-axis TDE jets. 
These authors proposed that  the late radio flare observed in AT2018hyz indicated an off-axis jet \cite{Matsumoto2023,Sfaradi2023}, opening the venue to identify relativistic jets that are pointing away from us.

\section{Hillas limits from late radio observations }
\label{sec:Hillas}
\subsection{The Hillas criterion}
Already in 1984, Hillas \cite{Hillas1984} 
compared the minimal acceleration time to the escape time and 
obtained a necessary condition for the maximal energy, $\cal E$, to which a source can accelerate a particle. In the simplest version of this condition, we have 
\begin{equation}
{\cal E} =  Z e B R \beta \Gamma  \ \ \ {\rm or } \ \ \ \ \frac{\cal E}{\rm eV} \approx 300 Z  \big(\frac{B}{\rm Gauss}\big) \big(\frac{R}{\rm cm}\big) \beta \Gamma\ , 
\label{eq:Hillas}
\end{equation}
where  B is the magnetic field (measured in the rest frame), R is the size of the source, $\beta$ and $\Gamma$ are the source velocity and its corresponding Lorentz factor and Z is the charge of the accelerated particle. 

Clearly the condition \ref{eq:Hillas} should be considered carefully within each model. For example the factor $\beta$ might not be relevant if the source is static. When considering jets, it is possible, depending on details of the acceleration process,  that the accelerated particles escape sideways \cite{Globus2023}. This shortens the escape time by a factor $\theta$, the opening angle of the jet. In this case, we have:
\begin{equation}
{{\cal \tilde  E}} =  Z e B R \beta \Gamma \theta \ \ \ {\rm or } \ \ \ \ \frac{{\cal \tilde E}}{\rm eV} \approx 300 Z  \big(\frac{B}{\rm Gauss}\big) \big(\frac{R}{\rm cm}\big) \beta \Gamma \theta \ .
\label{eq:Hillastheta}
\end{equation}
These estimates of the magnetic field and the size of the system imply associated luminosities of $L>8.4 \times 10^{44} Z^{-2} \Gamma^2 ({\cal \tilde E})/(10^{20} {\rm  eV})^2\mbox{erg s}^{-1}$ (e.g. \cite{Globus2023}). The sources are extremely powerful if UHECRs are protons. Less powerful, but still noticeable within local distances, if UHECRs are heavier nuclei.

\subsection{Late radio observations of TDEs}
\label{sec:hillastde}
Late radio observations enable us to identify jets whose prompt emission is not observed. By now about half a dozen TDEs have displayed late ratio flares. Among those, the rapid rise of the radio emission in AT2018hyz several years after the event \cite{Cendes2022,Horesh2022} suggests that this TDE harbored a powerful relativistic jet that was pointing away from us \cite{Matsumoto2023,Sfaradi2023}. 

In this case the late radio emission is generated by relativistic particles accelerated in shocks between the jet material and the surrounding circum-nuclear matter.  These 
observations are a calorimetric tool that enables us to measure the actual energy (not just the isotropic equivalent energy) of these jets. In some cases, they also enable a determination of the jet's opening angle.   With detailed modeling of the radio  emission, we can also reveal the conditions within these jets and explore whether they satisfy the Hillas criteria. 

In the following, we focus on the analysis of the radio emission from two of those jets: Sw J1644, the first discovered and best followed jetted TDE and AT2018hyz, which is an example of an ``off-axis" one in which prompt emission was not observed. We also consider the implications of the jet observed in AT2022cmc. 

{\bf Swift J1644} was the first identified jetted TDE. Its initial unique gamma-rays signal showed rapid variability, suggesting that it was a white dwarf that was disrupted \cite{Krolik2011}. The prompt non-thermal gamma-rays and X-ray from the  galactic nucleus \cite{Bloom2011,Burrows2011} were followed by a strong radio signal  \cite{Zauderer2011} confirming the original suggestion that the event involved a relativistic jet. 
The radio light curve was surprising as it increased during the first few hundred days. Equipartition analysis revealed that the  energy of the outflow that produced this radio signal increased by a factor of ten with time \cite{Barniol2013,Zauderer2013,Eftekhari2018}. While it was suggested that this is due to a structured jet with significant energy within wings \cite{Mimica2015,Generozov2017},  a simpler self-consistent explanation is that Sw J1644 was a slightly off-axis jet \cite{Beniamini2023}. Namely, while the jet width was $\sim 21^o$ our viewing angle was $\sim 30^o$. The observed radio signal increased as the relativistic jet's beam expanded sideways and more radiation was beamed towards us as the jet slowed down. 

The model developed in \cite{Beniamini2023} considers, for simplicity,  a relativistic `top-hat' jet with opening angle $\theta_0$ that has a uniform kinetic energy per unit solid angle. The jet decelerates due to its interaction with the external medium (assumed to have a radial profile $n\propto r^{-k}$) and its dynamics can be described analytically in two limiting cases, one in which the jet is assumed to spread sideways at a velocity of order the speed of sound (maximal lateral spreading) and the other in which the jet maintains its initial opening angle until it becomes Newtonian (no lateral spreading). 

The forward shock driven by the blast wave into the circum-nuclear matter amplifies the magnetic field and accelerates electrons to relativistic velocities, causing them to radiate their energy via synchrotron and synchrotron self-Compton. The temporal evolution of the flux at a given frequency band can then be calculated for an observer at an arbitrary viewing angle relative to the jet using the appropriate Doppler factor corrections. The model is uniquely described (at all times and frequencies) by a set of 9 parameters, three associated with the jet, $E_{\rm k,iso},\Gamma_0,\theta_0$ (the jets' isotropic equivalent kinetic energy, its initial Lorentz factor and angular width) three describe the microphysics, $p,\epsilon_{\rm B},\epsilon_{\rm e}$ (the power-law index of the electron's distribution, and the magnetic and electronic equipartition parameters), two characterize the surrounding environment, $k,n(R_0)$ (the power-law index of the external density and its normalization) and one identifies the viewing angle, $\theta_{\rm v}$. The multi-band observations can then be fit with this model to constrain these model parameters.

\begin{figure}[h]
\centering 
\includegraphics[width=.45\textwidth]{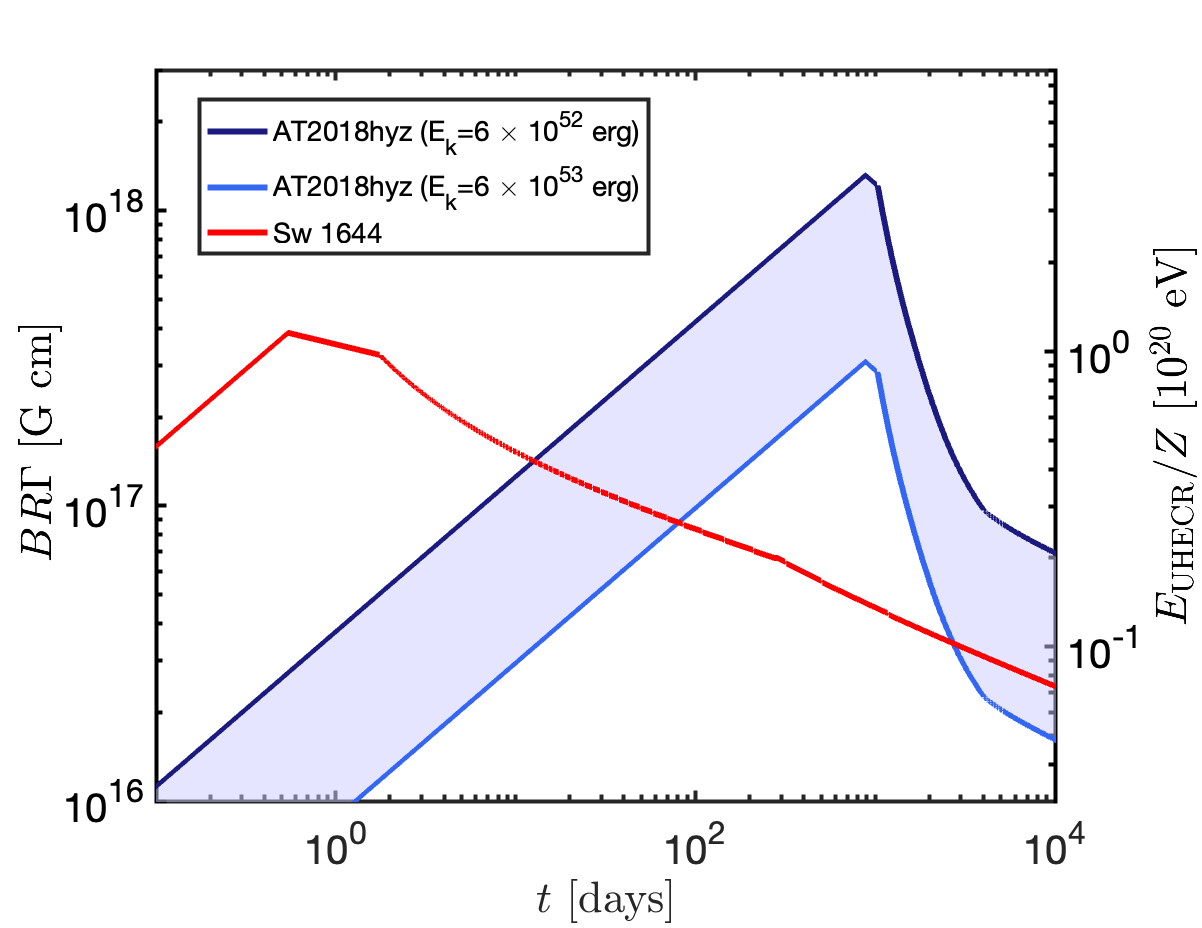}
\includegraphics[width=.45\textwidth]{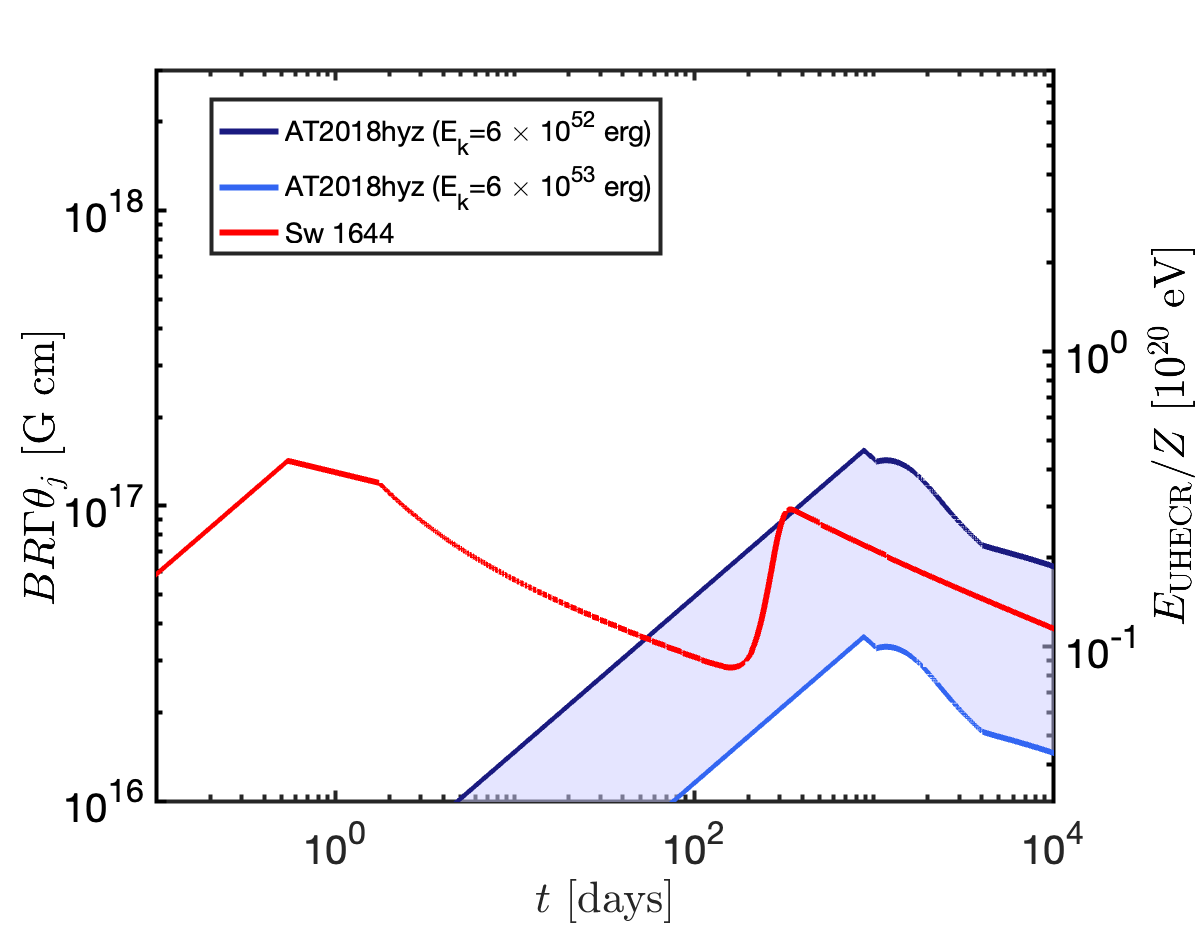}
\caption{The different Hillas criteria with (right) and without (left) particles being able to escape sideways (Eqs. \ref{eq:Hillas} and \ref{eq:Hillastheta}) respectively, for Sw J1644 and for AT2018hyz. For AT 2018hyz two estimates are shown, corresponding to different assumptions concerning the isotropic equivalent energy.}
\label{fig:BGammaR} 
\end{figure}

Remarkably, the conditions found within these shocks are suitable for accelerating UHECRs. Namely, these shocks are among the few astronomical sources that satisfy the Hillas criteria for accelerating particles up to $10^{20} eV$. In particular, Beniamini et al., \cite{Beniamini2023} modeled the light curve of Sw J1644 at several radio frequencies and estimated the emitting region's magnetic fields and size. This model gives us $B, R, \Gamma$ and $\theta$ as a function of time. 
Fig. \ref{fig:BGammaR} depicts the factor $B \Gamma R$ for the shocks in Sw J1644. This result confirms earlier findings \cite{Farrar2014} that the shock between the jet and the circum-nucleus gas of the host galaxy satisfies the Hillas criteria (with $E=10^{20}$ eV even for protons). 

As mentioned earlier, it is not clear whether the accelerated particles can escape sideways from the jet. If they do, the Hillas condition is modified (see Eq. \ref{eq:Hillastheta}). The right panel of Fig. \ref{fig:BGammaR} depicts $B\Gamma R \theta$ the Hillas condition when allowing for sideways escape of accelerated particles. In this case, the values are lower and the shock cannot accelerate protons to the highest observed energies, $\sim 10^{20}$ eV. However, it is capable of accelerating heavier nuclei to this energy. 
When the jet angle is taken into account, the factor $B R \Gamma \theta$ increases sharply at around 300 days. {This is because the best-fit model for this event involves a jet that isn't spreading significantly during the relativistic phase. The late rise at 300 days, is approximately when the jet becomes Newtonian and starts spreading sideways quickly.}

{\bf AT2018hyz} was an optically detected TDE located at $z=0.0457$. Its prompt signature didn't show non-thermal emission corresponding to  a jet and early radio signal was not detected. A surprising rapidly rising radio flare was detected from AT 2018hyz 972 days after optical discovery  \cite{Cendes2022,Horesh2022}. The initial interpretation was of a delayed launching of a jet \cite{Cendes2022}. However, following an off-axis equipartition analysis \cite{Matsumoto2023}, Sfaradi et al. \cite{Sfaradi2023} have recently shown that an off-axis jet launched at the time of the optical discovery was a natural self-consistent interpretation of the radio data. 
We didn't see the prompt emission in AT2018hyz  as it is beamed away from us. As the shock decelerated, it moved into our line of sight. At the same time, it accumulated more and more emitting electrons. It is for this reason that the radio signal increased so rapidly. 
This interpretation implies that the jet in AT2018hyz was very powerful with $E_k \ge  3 \times  10^{52} $ erg and an opening angle of $\sim 7^o$. The jet that was launched at the time of the stellar disruption (or shortly after that) was at $\sim 42^o$ to our line of sight. 

The magnetic field in AT2018hyz is uncertain.  This arises since, for the best-fit model for AT2018hyz, most of the data corresponds to a single spectral band of the synchrotron spectrum - leading to a well-understood degeneracy between magnetic field, isotropic equivalent energy and external density (see \cite{Sfaradi2023} for details). Still, the magnetic field can be bounded from below. A lower magnetic field (a lower $\epsilon_B$) requires a larger total energy. Considering two possible values of the total energy $6 \times 10^{52}$ erg and $6 \times 10^{53}$ erg (the latter can be regarded as a safe upper limit), we present in Fig. \ref{fig:BGammaR} the two possible estimates for the Hillas conditions Eq. \ref{eq:Hillas}  and \ref{eq:Hillastheta}. Like in Sw J1644, if sideways escape of the accelerated particles is negligible, the shocks in AT2018hyz can accelerate even protons to $10^{20}$eV. If sideways escape is significant, the shocks can still accelerate nuclei (with $Z\ge 10$) to this energy, even when considering the lower values for the magnetic field. 

Fig. \ref{fig:BGammaR} shows that for off-axis jets, the optimal conditions for UHECR acceleration are obtained when the jet comes into view of the observer ($\Gamma \approx 1/\theta_{\rm v}$ for far off-axis jets). As such, there could be a delay of several hundred days between the optical TDE and the peak of UHECR production from such sources. This delay is irrelevant as UHECR diffusion between the source and Earth results in much longer delays. However, it is important when considering the association with possible secondary neutrinos. 

{\bf AT2022cmc} was a third jetted TDE  \cite{Andreoni2022}. It was a rather distant event, located at $z=1.19$.  Its implied isotropic equivalent energy was $10^{53}-10^{54}$ erg. If viewed on-axis, there is no clear estimate of the opening angle. However, the 15 GHz radio light curve during the first 100 days (in the rest frame) resembles the one of Sw J1644 \cite{Rhodes2023}. Lacking further data, we note the possibility that this is also a slightly off-axis jet but adopt in the rest of the text, the original interpretation \cite{Andreoni2022}, that it was viewed on-axis. 

\section{Rates and total emitted power density}
\label{sec:rates}

\subsection{Rates}
The overall rate of optical TDEs is estimated as $\mathcal{R}\approx 10^{-6}\mbox{ Mpc}^{-3} \mbox{ yr}^{-1}$ \cite{Stone2020rate}. However, the question what is the rate of jetted TDEs is much more subtle as those are fewer and detection selection effects are more dubious.  

For a given jetted TDE event, we can use its distance (corresponding to a cosmological redshift $z_{\rm ob}$ and volume $V_{\rm ob}$ within which there is at least one event), the observation time during which similar events could have been detected ($T_{\rm ob}$) and the sky coverage fraction ($f_{\rm sky}$) of the relevant telescope/s to estimate the intrinsic rate of such events. Since events generally get dimmer with increasing viewing angle, this estimate can be considered as an approximate estimate of similar events with $\theta<\theta_{\rm v}$, (where $\theta_{\rm v}$ is the viewing angle to the event considered). Therefore, we can write 
\begin{equation}
\label{eq:Ratej}
     \mathcal{R}_j\approx 0.7\frac{(1+z_{\rm ob})}{f_{\rm sky}V_{\rm ob} T_{\rm ob} (1-\cos \theta_{\rm v})}
\end{equation}
where we have assumed that intrinsically the events are double-sided jets (such that $\theta_{\rm v}<\pi/2$) and that the intrinsic rate of events does not strongly evolve with redshift (TDEs are observed from relatively low redshifts). For clarity, we have also approximated $\int^{z'} (dV/dz)(1+z)^{-1}dz\approx V(z')/(1+z')$ which is a good approximation for the range of redshifts from which  TDEs are observed.

Applying the estimate given by Eq. \ref{eq:Ratej} to Sw J1644 (with $z_{\rm ob}=0.35$, $T_{\rm ob}\approx T_{Swift}\approx 20\mbox{ yr}$, $f_{\rm sky}=f_{\it Swift}\approx 0.1$, $\theta_{\rm v}\approx 0.5$), we find $\mathcal{R}_{1644}\approx 3\times 10^{-10} \mbox{ Mpc}^{-3} \mbox{ yr}^{-1}$. During this time {\it Swift}-BAT  detected two other TDE jet candidates, Sw J2058 and Sw J1112 \cite{Cenko2012,Brown2015}. However, since the distance to those events was significantly larger than that of Sw J1644, they are sub-dominant in their contribution to the estimated rate.

The estimate given by Eq. \ref{eq:Ratej} can also be applied to radio-monitored optically-discovered TDE jet candidates such as AT2018hyz. Taking $z_{\rm ob}=0.045$, $T_{\rm ob}\approx T_{\rm ZTF}\approx 5\mbox{ yr}$, $f_{\rm sky}=f_{\rm ZTF}\approx 0.6$, $\theta_{\rm v}\approx 0.85$ we get $\mathcal{R}_{hyz}\gtrsim 2\times 10^{-8}\mbox{ Mpc}^{-3}\mbox{ yr}^{-1}$. This is very likely a lower limit, considering that it requires an optical TDE candidate to have long-duration radio follow-up observations.

AT2022cmc provides another constraint on the rate of jetted TDEs. The estimated  observed rate of AT2022cmc-like events with ZTF is $\mathcal{R}_{cmc}\approx 2\times 10^{-11}(1-\cos \theta_{\rm v})^{-1}\mbox{ Mpc}^{-3}\mbox{ yr}^{-1}$  \cite{Andreoni2022}. AT2022cmc's 100GHz light curve is monotonically declining in time from the earliest observations at $t\gtrsim 10$\,days after its onset. This motivated  modeling it   as being viewed on-axis  \cite{Andreoni2022}
(see also \cite{2023MNRAS.522.4028M}). More recently,  observations at 5-30 GHz  \cite{Rhodes2023}  show a slowly rising light curve between 5-100 days after onset. Such a rise could, in principle, be attributed to an observed band being below the synchrotron injection frequency at the time of observations (see Fig. 7 of \cite{Beniamini2023}). Comparison of  their results to the lightcurve of Sw 1644, suggests that AT2022cmc is consistent with being viewed slightly off-axis.  
Considering an opening angle $\theta_{\rm j}\approx 0.1$ and an on-axis viewing angle, we find that $\mathcal{R}_{cmc}\approx 4\times 10^{-9}\theta_{\rm j,-1}^{-2}\mbox{ Mpc}^{-3}\mbox{ yr}^{-1}$. 
This estimate would increase  if the event were observed slightly off-axis, as the radio afterglow suggests. Further radio data is required for a detailed analysis.

We find  three different estimates of the rate, ranging from $3 \times 10^{-10}$Mpc$^{-3}$ yr$^{-1}$ to $2 \times 10^{-8}$Mpc$^{-3}$ yr$^{-1}$.
Namely, a fraction of $10^{-4}$ to $10^{-2}$ of jetted out of total TDEs. The large range is surprising, even given the small number statistics involved. It may indicate a somewhat different nature of  the three TDEs. The least certain of these is the rate associated with AT2022cmc, which depends on an unknown beaming factor and could end up being comparable to either of the other two rates. Furthermore, it is possible that AT2018hyz-like events are indeed much more common than Sw J1644-like events. This is because we have no detections of early X-ray and $\gamma$-ray emission from AT2018hyz. If similar jets, as viewed by far off-axis observers, are initially extremely dim in these bands, this would explain why they could be lacking from the {\it Swift} sample despite being intrinsically common. Finally, it is worth considering that if AT2018hyz was an on-axis event, as suggested by \cite{Cendes2022}, the rate implied from Eq. \ref{eq:Ratej} (and hence the apparent discrepancy with Sw J1644) would have been even greater, by perhaps two orders of magnitude. 

\subsection{Power density estimates}

Using the rates above, we can estimate the energy input  per unit volume and time from a population of jets for which these events are representative. The best fit model for Sw J1644 and AT2018hyz, suggest collimated-corrected kinetic energy of $E_{\rm k,1644}=7\times 10^{52}\mbox{ erg}$
\cite{Beniamini2023}, $E_{\rm k,hyz}=6.5\times 10^{52}\mbox{ erg}$ \cite{Sfaradi2023} 
\footnote{Cendes et al. \cite{Cendes2022} suggested that AT2018hyz was an on-axis outflow with late time energy injection. The minimal electron and  magnetic energy they estimate, using equipartition analysis, is $\sim 6\times 10^{49}$\mbox{ erg}. The true kinetic energy of the outflow in an energy injection scenario is typically significantly larger considering: (i) A factor of $\sim 10$ conversion from electron and magnetic energy to kinetic energy; (ii)   The equipartition analysis does not account for a self-consistent evolution of the spectral frequencies and peak flux (see \cite{Beniamini2023} and \cite{Sfaradi2023} for details) and as such it under-predicts the required energy (for example by not accounting for the fact that the synchrotron spectrum is likely deep in the slow cooling regime, and the radiated energy is therefore much smaller than the energy stored in the electrons).  Since in both the late energy injection and the off-axis interpretation, the peak of AT2018hyz is obtained at a  mildly relativistic stage, the energy required for the two scenarios should be similar. This means that the volumetric energy injection rate would be larger by one over the beaming factor for the late energy injection scenario, which could correspond to an overall estimate of the energy injection rate greater by up to two orders of magnitude.}. For AT2022cmc, we use the kinetic energy estimate of \cite{Andreoni2022}, $E_{\rm k,cmc}\approx 1.5\times 10^{51}\theta_{\rm j,-1}^2\mbox{ erg}$.
Putting this together we have:
\begin{eqnarray}
    E_{\rm k}\mathcal{R}\approx\left\{ \begin{array}{ll}2.2\times 10^{43}\mbox{erg Mpc}^{-3}\mbox{ yr}^{-1} & \mbox{Sw J1644}\ ,\\
1.4\times 10^{45}\mbox{erg Mpc}^{-3}\mbox{ yr}^{-1} & \mbox{AT2018hyz}\ ,\\
6\times 10^{42}\mbox{erg Mpc}^{-3}\mbox{ yr}^{-1} & \mbox{AT2022cmc}.
\end{array} \right.
\end{eqnarray}
The results are summarized in table \ref{tab:table}.

\begin{table}[h]
\caption{\label{tab:table}%
A comparison of rates and energy input rate of the three different TDEs }
\begin{tabular}{|c||c|c|c|}
\hline
\textrm{TDE}&
\textrm{Sw J1644}$^a$&
\textrm{AT2018hyz}$^b$&
\textrm{AT2022cmc}$^c$\\
\hline
\hline
\textrm{Red Shift} & 0.35 & 0.0457 &  1.19\\
\hline
\textrm{Kinetic energy} (erg) &$ 7 \times 10^{52}$ &$ 6.5 \times 10^{52}  $&  $1.5 \times 10^{51}$~$^{d,e}$ \\
\hline
\textrm{Rate} ({Mpc}$^{-3}$yr$^{-1}$)& $3 \times 10^{-10}$ & $2 \times 10^{-8}$  & $4 \times 10^{-9}$~$^{d}$ \\
\hline
\textrm{Energy input rate} (erg~{Mpc}$^{-3}$yr$^{-1}$) &$2.2\times 10^{43}$ & $1.4\times 10^{45}$~$^{f}$ & {$6\times 10^{42}$~$^{d,e}$} \\
\hline
\end{tabular}
\hfill\break \\
a. {From \cite{Beniamini2023}.  }
b. {From \cite{Sfaradi2023}.  }
c. {From \cite{Andreoni2022}, (see also \cite{2023MNRAS.522.4028M}).  }
d. The rate and the kinetic energy depend on the assumed jet opening angle. However, the energy input rate doesn't depend on it, if the jet was viewed on-axis. 
e. The energy and energy input rate will be slightly larger if the event was observed slightly off-axis. f. The rate and the overall energy injection rate will be larger by up to two orders of magnitude if interpreted within the late jet scenario of \cite{Cendes2022}.
\end{table}

\section{Conclusions}
\label{sec:conc}
Detailed modeling of TDEs radio emission have shown that  jetted TDEs are capable of accelerating UHECRs nuclei to the highest observed energies. Depending on the nature of the acceleration process, these jets may even be able to accelerate protons (if there is no sideways escape from the jet). These recent results confirm the earlier findings \cite{Farrar2014} on this question.

The situation is more complicated concerning the luminosity density. Here, the estimates based on Sw J1644 are about an order of magnitude below ${\cal L}_{\rm UHECR}$, and an even lower estimate is obtained from AT2022cmc. However, At2018hyz sheds a different light on this issue. While the lower limit estimate for its kinetic energy is comparable to the kinetic energy estimate of Sw J1644, because of its small redshift its implied event rate hundred times larger. With these parameters the estimated luminosity density, $1.4 \times 10^{45}$ erg {Mpc}$^{-3}$yr$^{-1}$ is sufficient to produce the observed UHECR flux, making jetted TDEs a viable candidate for UHECR sources candidate. 

It is interesting to compare jetted TDEs and GRBs as potential UHECR sources. Remarkably in both cases, the acceleration model is rather similar. The shocks produced in the interaction of the jetted TDEs with the surrounding matter are similar to GRBs afterglows' shocks\footnote{Within GRBs UHECR acceleration can also take place within the internal shocks.} (note that the GRB jets have a higher Lorentz factor). Both events have a comparable rate. However, it seems that the jetted TDEs are more energetic, making their total energy input comparable to the required ${\cal L}_{\rm UHECR}$.


As TDEs are transient and as diffusion of UHECRs prolongs their travel from sources to Earth and also diverts them from their original directions, we cannot expect a direct identification of an excess of UHECRs with a TDE. However, indirect evidence could arise from the detection of a secondary high-energy neutrino that could be produced by an interaction of a UHECR with the ambient matter of radiation field. Interestingly, some evidence for an association of high-energy neutrinos with TDEs has already been found \cite{Stein2021,Reusch2022}. { In all cases, the arrival direction of the neutrinos was consistent with that of the TDE, and the former were detected with a time delay of order $\sim 200$ d from the latter. The three TDEs associated with neutrino emission, have been observed with IR dust echoes which overlapped in time with the neutrino detections. It has been suggested that this dust echo may be the photon target for the cosmic rays to produce the neutrinos \cite{Winter2023}, which would require the primaries to be in the UHECR range. Since neutrino produced by the interaction of IR photons with the cosmic rays, preferentially move along the direction of the UHECR primary, this explanation, requires the jets in those events to be moving towards the observer, or at most to be oriented mildly off-axis from the observer. For a mildly off-axis, depending on the exact jet orientation and on the location of the dust cloud relative to the black hole, the neutrino emission in such a scenario may appear after or even before the radio peak associated with the jetted TDE. TDE jets that are significantly off-axis may produce IR echos but we won't be able to observer their neutrinos. }

To conclude, we have shown here that TDE jets satisfy the Hillas criteria. Evidence from an off-axis jet observed in AT2018hyz suggest that   these jets  also carry sufficient power, { depending, of course, on the efficiency of UHECR acceleration, } to produce the whole flux of UHECRs above 5 EeV.
As new data on TDEs are accumulating rapidly now and with the realization that we can identify jetted once via their late radio flares we can expect to rule out or support this model by further observations of other TDEs that will improve the current poor statistics.

\acknowledgments

We acknowledge support from an Advanced ERC grant MultiJets (TP) and  grant no. 2020747 from the United States-Israel Binational Science Foundation (PB).

\section*{Data Availability}

No new data was generated in this research. Details about modeling the light curve of off-axis TDE are given in \cite{Beniamini2023} and \cite{Sfaradi2023}.

\bibliographystyle{ieeetr}



\end{document}